\documentclass[conference]{IEEEtran}
\IEEEoverridecommandlockouts
\usepackage{cite}
\usepackage{amsmath,amssymb,amsfonts}
\usepackage{algorithmic}
\usepackage{graphicx}
\usepackage{textcomp}
\usepackage{xcolor}
\usepackage{bm}      
\usepackage{algorithm}
\usepackage{booktabs}
\usepackage{multirow}
\usepackage{caption}

\def\BibTeX{{\rm B\kern-.05em{\sc i\kern-.025em b}\kern-.08em
		T\kern-.1667em\lower.7ex\hbox{E}\kern-.125emX}}
\makeatletter
\newcommand{\linebreakand}{%

\end{@IEEEauthorhalign}
	\vspace{-8mm}
\hfill\mbox{}\par
\mbox{}\hfill\begin{@IEEEauthorhalign}
}
\makeatother

\begin{document}
	
	\title{PatSTEG: Modeling Formation Dynamics of Patent Citation Networks via The Semantic-Topological Evolutionary Graph}

	\author{
		\IEEEauthorblockN{Ran Miao$^{1}$, Xueyu Chen$^{1}$, Liang Hu$^{1,3,\dagger}$, Zhifei Zhang$^{1}$, Minghua Wan$^{2}$, Qi Zhang$^{1,3}$, Cairong Zhao$^{1}$}
		\IEEEauthorblockA{$^1$\textit{Tongji Unversity, Shanghai}\quad $^2$\textit{Nanjing Audit University, Nanjing}\quad $^3$\textit{DeepBlue Academy of Science, Shanghai}\\
			\{2233054, lianghu, zhifeizhang, zhangqi\_cs, zhaocairong\}@tongji.edu.cn, cxya@163.com, wmh36@nau.edu.cn}

		
	}

	\maketitle
	\begin{abstract}
         \renewcommand{\thefootnote}{}
        \footnotetext{$^{\dagger}$ Corresponding author.}
		Patent documents in the patent database (PatDB) are crucial for research, development, and innovation as they contain valuable technical information. However, PatDB presents a multifaceted challenge in comparison to publicly available preprocessed databases due to the intricate nature of the patent text and the inherent sparsity within the patent citation network. Although patent text analysis and citation analysis bring new opportunities to explore patent data mining, no existing work exploits the complementation of them. To this end, we propose a joint semantic-topological evolutionary graph learning approach (PatSTEG) to model the formation dynamics of patent citation networks. More specifically, we first create a real-world dataset of Chinese patents named CNPat and leverage its patent texts and citations to construct a patent citation network. Then, PatSTEG is modeled to study the evolutionary dynamics of patent citation formation by considering the semantic and topological information jointly. Extensive experiments are conducted on both CNPat and public datasets to prove the superiority of PatSTEG over other state-of-the-art methods. All the results provide valuable references for patent literature research and technical exploration.
	\end{abstract}
	
	\begin{IEEEkeywords}
		patent analysis, patent citation network, semantic-topological
		evolutionary graph
	\end{IEEEkeywords}
	
	\section{Introduction}
	
	Patents are crucial assets that support innovation across a range of sectors and safeguard intellectual property rights~\cite{zhang2015patent}. Patent citation networks, which record the connections between patents via citations, provide insightful data on the interconnectedness, impact, and flow of information within certain technology sectors. In recent years, the method of graph theory has been widely used in different fields, such as social relations, information dissemination, citation relations and so on.~\cite{cui2018survey}. For example, Chen et al. ~\cite{chen2019link} proposed Structural Deep Network Embedding, which has a better performance on the BlogCatalog dataset than Node2Vec~\cite{grover2016node2vec}. Chakraborty et al. ~\cite{chakraborty2020patent} employed Exponential Random Graph Models (ERGMs) to analyze the citation networks. Chen et al. ~\cite{ma2022organization} presents a future-oriented framework based on the link prediction methods to research Alzheimer’s disease-related patents. Ji et al.~\cite{ji2019patent} propose a sequence-to-sequence model which employs an attention-of-attention mechanism. 
	
	\begin{figure}[htbp]
		\vspace{-5mm}
		\centering
		\resizebox{\linewidth}{!}{
			\includegraphics[width=\linewidth]{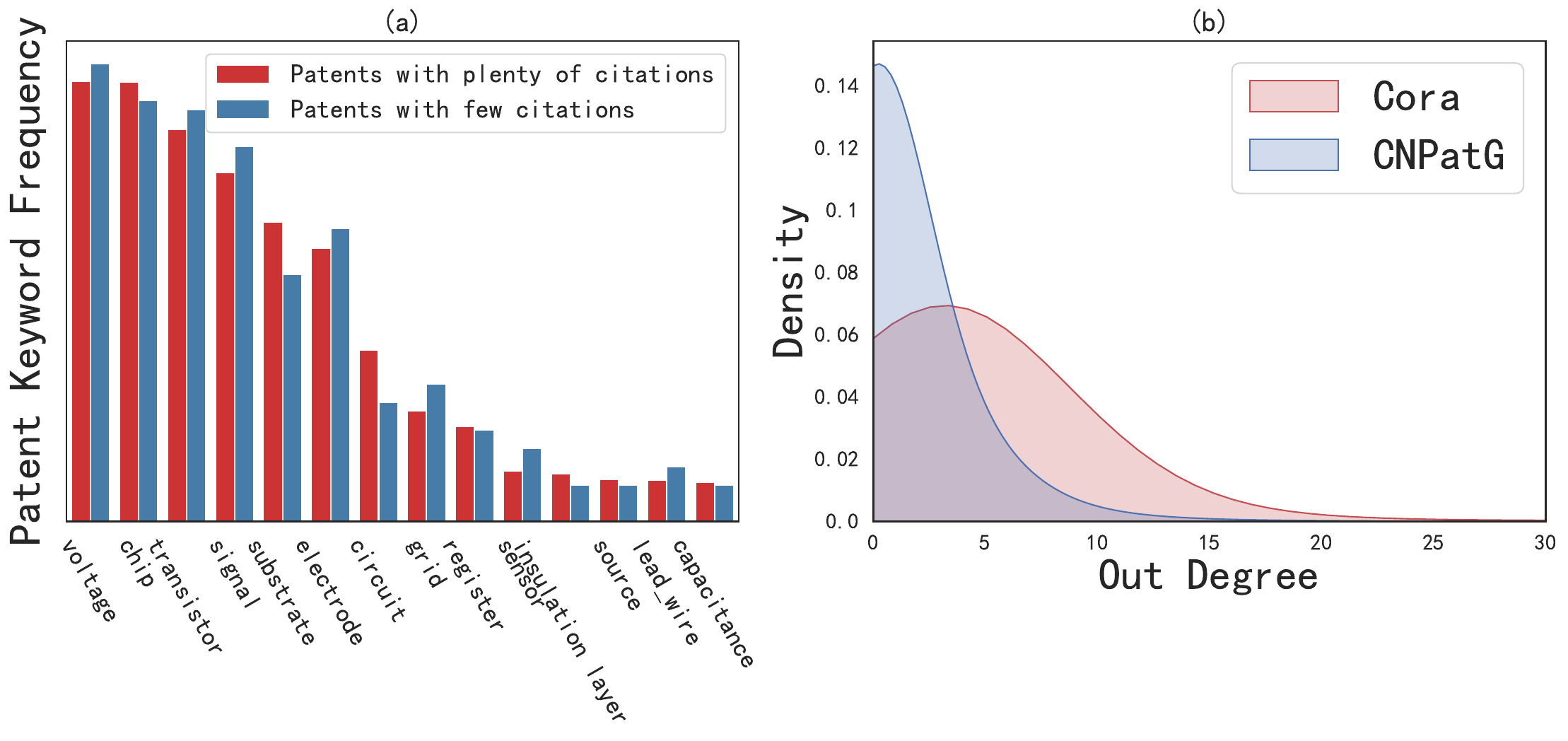}
		}
		\caption{(a) The similar keyword distributions of patents both related to chip technology, separately from patents with plenty of citations and patents with few citations. (b) The out-degree distribution of CNPatG is more imbalanced than Cora. }
		\label{fig:degree}
		\vspace{-4mm}
	\end{figure}
	
	However, these methods usually learn the topology of the network, which brings a problem in that they cannot learn the rich information (such as text) contained by nodes in the network~\cite{gao2018deep}. Current research on link prediction in patent citation networks typically focuses on the graph's \textbf{topological structure}, while it often overlooks the valuable \textbf{textual attributes}. Because they ignore an important fact that there is a strong coupling between the relevant information contained in the patent and that the textual content and other features of the patent determine the citation of the patent. 
	
	Our analysis reveals a consistent similarity in keyword distributions among patents related to chip technology, irrespective of whether they receive ample citations or minimal citations (Fig.~\ref{fig:degree} a). This observation underscores the heightened complexity inherent in the study of the CNPatG dataset relative to conventional databases. It also emphasizes the critical role of textual content analysis, particularly in scenarios characterized by a paucity of citations. Furthermore, as illustrated in Fig.~\ref{fig:degree}, it is evident that the out-degree distribution of the CNPatG dataset (Fig.~\ref{fig:degree} b) exhibits a significantly higher degree of imbalance when compared to the Cora dataset. Moreover, current research does not focus on the \textbf{multi-aspect of patent citations}. A patent can be cited from many patents for different reasons, since it may have several technology topics. It is significant for patent researchers to identify citations of different topics.
	
	Therefore, the deficiencies in link prediction in patent citation networks are obvious and the task of link prediction in such networks continues to pose several challenges. (1) As new patents are filed and published, it takes time for citations to be made. Consequently, the absence of citations only represents the relationship between patents that have not yet been discovered. (2) The enormous number of patents and the complex nature of their interconnections pose significant challenges in comprehensively analyzing the entire network manually. As shown in Fig.~\ref{fig:degree}, we compare the out-degree distribution of the PubMed data set and CNPatG. It is obvious that most patent citations are small, and only a few patent citations are large. (3) It is necessary to recognize which topic the citation means. For example, two patents cite from the same patent which composites two kinds of technology. However, these two patents can relate to the two different technology topics separately. Distinguishing citations of different topic aspects is significant for link prediction in the citation network. Therefore, there is a compelling need for efficient methods to predict future or absent citations in patent citation networks.

	For those drawbacks and challenges in research, our motivation is to combine the topological structure of citation networks with the textual attributes of patents for link prediction, while also enabling the mining of patent citations for different topics. Furthermore, our model mines patent citations across diverse topics and offers visualizations for multi-aspect citations, which greatly benefits patent researchers.

	

	The emergence of new patent nodes and new citation relationships makes the citation network expand continuously as time goes on. The multi-aspect citation links also make the network more complicated. Patents contain a lot of information, which promotes the formation of citations between patents and reveals the law of technological development. In this paper, we define it as the link prediction problem in the citation network and discuss and explain the link prediction. Setting patents as nodes, a dynamic patent citation network is constructed to analyze patents. The main contributions of this paper are as follows:
	
		(1) We utilize Intellectual Property Publishing House's patents to create CNPat data sets, including 8 technical fields categorized by IPC codes from A to H.
		
		(2) A new citation network analysis model (PatSTEG) is designed. By analyzing the characteristics of the data in the patent database, we construct a patent citation network and use the idea of dynamics to learn the embedded representation of patent data, which is used to predict and explain the citation relationship between patents.
		
		(3) We have proved the feasibility of PatSTEG. A large number of experiments on real-world Chinese patent databases and benchmark databases show that PatSTEG outperforms baselines in link prediction.
	

	\section{Related Works}
	
	\subsection{Citation Analysis}
	
	
	The patent citation relationship reflects the influence of patents and the development of technology. Many studies use citations to analyze scientific works. Patent citation analysis began to gain attention in the 1990s. Many subsequent studies have shown that the number of patent citations is positively correlated with patent value, and the number of patent citations is also the most commonly used independent test index. Although the forward reference has traditionally been used as an analysis index of patents, its practicability is mainly retrospective, such as collecting the reference count of public patents. For example, Chakraborty et al.~\cite{chakraborty2020patent} used Exponential Random Graph Models (ERGM) to analyze the citation network from a statistical point of view. It concluded that the same application countries and writing languages promote the citation of patents, and recent patents are more likely to be cited.
	
	
	\subsection{Attributed Network Representation}
	Many real-world problems can be represented by graphs, such as social networks, technology prediction, etc.~\cite{2016Scalable,2020Collaboration,abs-2309-09179,NaseemTZHN23}. Traditionally, most problems are designed as static graphs. However, the nodes, attributes, and edges of many applications will change over time, and the static graph cannot represent the continuous changes of the network. Relevant work has been done to study the representation methods of dynamic networks, mainly including discrete networks and continuous networks~\cite{2021Foundations}. The discrete network uses multiple network snapshots to represent the graph structure information of different times in the dynamic network. The static graph analysis method is used on each snapshot~\cite{2018Dynamic}. The continuous network preserves the dynamic nature of the network, incorporating accurate time information into its learned graph representation. The continuous network representation is also more complex, as discussed in \cite{xu2020inductive} and \cite{2018Streaming}.
	
	\subsection{Textual Content Analysis}
	
	The value of the patent comes from the patent content. Besides the structural information available immediately after the publication of the patent, the text content of the patent (title, abstract, claim, etc.) also contains a lot of rich information~\cite{arts2021natural}. A classic patent text analysis method is to summarize and express patents by extracting key phrases or technical phrases in patents~\cite{liu2020technical}. To further analyze the semantic relationship between words or sentences, many works use the traditional semantic analysis methods, such as Latent Dirichlet Allocation (LDA) and Latent Semantic Analysis (LSA)~\cite{zhang2021integrating,ZhuZCA20,du2021personalized} or deep learning methods in the field of NLP to analyze patent texts. Denter et al. proposed text-based link prediction to predict the future bigrams, believing that promising patents contain a large number of future doublewords~\cite{denter2022forecasting}.
	
	\section{Problem Formulation}
	
	In this section, we will describe the patent citation network and formulate the link prediction problem in the patent citation network with auxiliary attributes. In this paper, we explore the ``Title", ``Abstract" and ``Claim" alone or jointly as side attributes. One advantage of this is that each patent contains text attributes, which means that PatSTEG is extensible. In addition, a large number of key technical phrases contained in these texts are one of the most important attributes of patents. Finally, a dynamic patent citation network with attributes is established to analyze the reasons for the formation of patent citations and predict possible future citations.
	
	\begin{figure}
		\centering
		\includegraphics[width=1\linewidth]{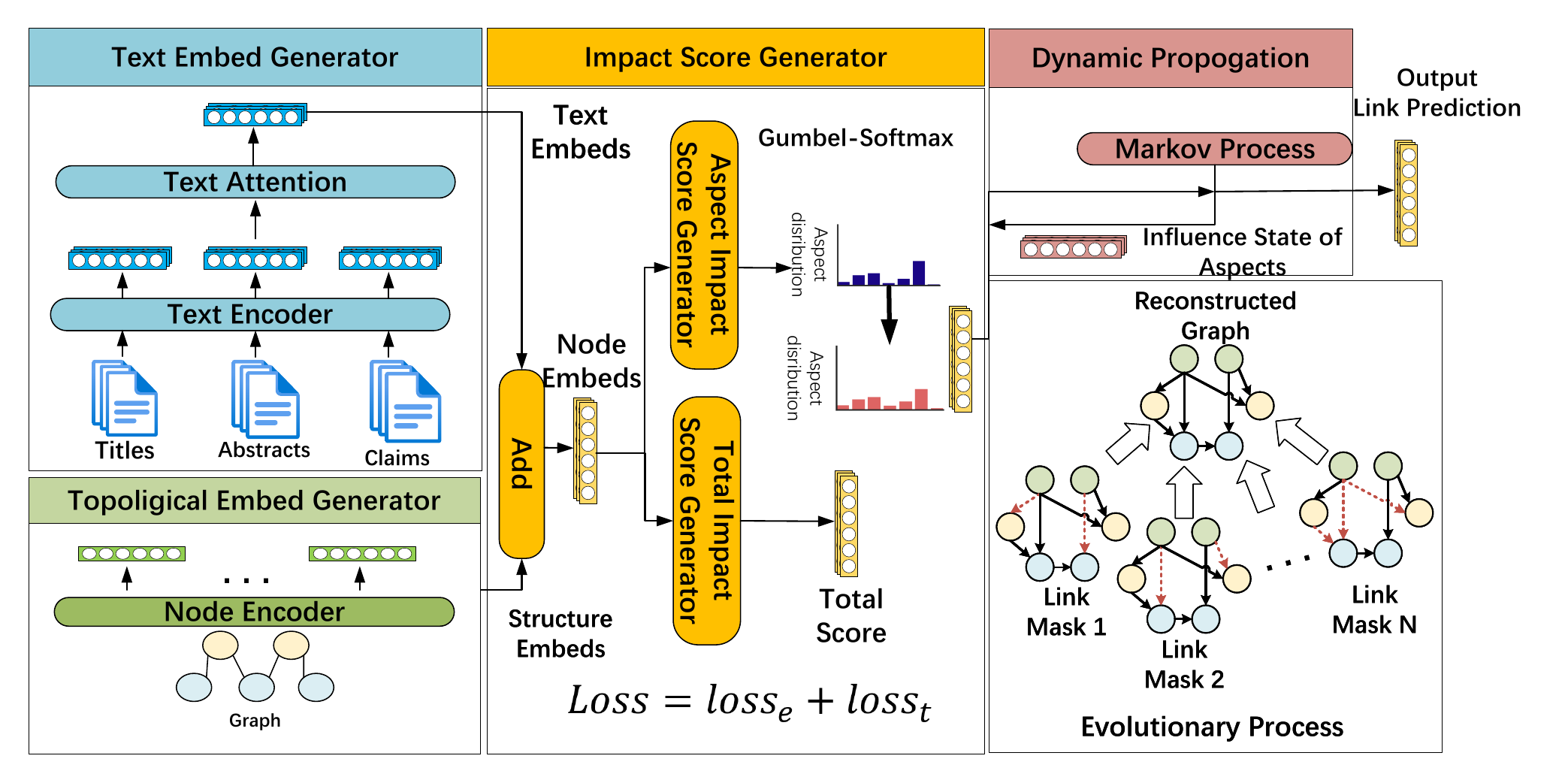}
		\caption{The architecture of the PatSTEG model with the evolutionary learning process. }
		\label{fig:architecture}
		\vspace{-4mm}
	\end{figure}
	
	\section{Methodology}
	To design an effective method for link prediction in patent citation networks, we propose a novel framework named PatSTEG. In the rest of this section, we introduce the proposed model in detail to show the formation process and evolutionary learning strategy of the network, which will be used for mining patent attributes.
	
	\subsection{Citation Network Construction}
	
	According to the analysis in Section 3.1, the topological structure of the patent citation network is highly sparse. In this case, it is necessary to use auxiliary attributes as a supplementary information source. The abundant text information contained in patents reflects the major technologies and patent categories, and it is precisely this information that captures the value of patents.
	
	
	%
	
	\subsection{Model Overview}
	
	Our goal is to learn the final patent citation network representation by dynamically learning each time snapshot. For the citation graph ${\tilde G}$, we establish two coupling systems ${S_Y}$ and ${S_D}$, which are initially expressed as Equation (1) and Equation (2). The architecture of the proposed PatSTEG model is illustrated in Fig.~\ref{fig:architecture}.
	\vspace{-2mm}
	\begin{equation}
		\\S_Y:\mathbf{Y^{(h+1)}} = {f_Y}(\mathbf{D^{(h)}},\mathbf{{Y^{(h+1)}}}\left| {\tilde G} \right.)
		\vspace{-2mm}
	\end{equation}
	\begin{equation}
		\\S_D:\mathbf{D^{(h+1)}} = {f_D}(\mathbf{Y^{(h)}},\mathbf{D^{(h)}}\left| {\tilde G} \right.)
		\label{S_Y_eq}
		\vspace{-2mm}
	\end{equation}
	
	${S_Y}$ aims to learn the citation linkage strength between each pair of patents. ${\mathbf{Y^{(h)}}}$ is the patent linkage score of influence in different aspects in the initial state at the $h$-th iteration, and ${\mathbf{Y}}$ is obtained through ${S_Y}$ learning; ${S_D}$ aims to update the influence status of different aspects of patents, and ${\mathbf{D^{(h)}}}$ is a new network structure after ${S_Y}$ learning at the $h$-th iteration.
	
	For each patent ${i}$ in the network, we propose to learn representation ${\mathbf{r}_i \in {\mathbb R^{L \times 1}}}$ of each node ${\mathbf{n}_i}$. To comprehensively represent the effective attributes in the patent. In addition to the topology information, we also consider the text information of the patent, for example, the title text.
	\vspace{-2mm}
	\begin{equation}
		\\\mathbf{r}_i = {l_2}\;(\mathbf{t_i} \oplus \mathbf{n_i})
		\label{r_i_eq}
		\vspace{-1.5mm}
	\end{equation}
 where ${l_2}$ denotes the ${L_2}$-norm, and $\mathbf{t_i}$ denotes the word vector representation of the text of patent ${i}$. We extracted ``Title", ``Abstract" and ``Claim" from the patent text respectively. Word segmentation is carried out on the extracted text for Chinese patents. Finally, word vector representations are learned by the GloVe algorithm~\cite{pennington2014glove}.
	
	Further, we analyze the node features extracted above. Suppose an edge between source node ${i}$ and target node ${j}$. As a rule, the citation between patents is determined by the impact of the target patent. We define $\mathbf{c}_{ij}$ to indicate citation effect from patent $j$ to patent $i$,
	
	\vspace{-2mm}
	\begin{equation}
		\\\mathbf{F}_{ij} = \sum {\mathbf{c}_{ij}}  + \sum {\mathbf{e}_{ij}}
		\label{F_ij_eq}
	\end{equation}
	\vspace{-4mm}
	\begin{equation}
		\\\mathbf{c}_{ij} = \mathbf{T}^d{d}_j
		\label{c_ij_eq}
	\end{equation}
	\vspace{-5mm}
	\begin{equation}
		\\\mathbf{e}_{ij} = \mathbf{r}_i \odot \mathbf{r}_j
		\label{e_ij_eq}
		\vspace{-2mm}
	\end{equation}
	where $\mathbf{F}_{ij}$ is used to indicate the total impact of patent citation between patent $i$ and $j$. It is summed over $\mathbf{c}_{ij}$ and $\mathbf{e}_{ij}$. ${\mathbf{T}^d \in {\mathbb R^{I \times I}}}$ is the weight matrix , and  ${\mathbf{d}_j \in {[0,1]^I}}$ denotes a state vector with ${I}$ aspects. $\mathbf{r}_i$ and $\mathbf{r}_j$ denote the representations of patent node $i$ and $j$, and ${\odot }$ represents Hadamar product. $\mathbf{e}_{ij}$ to represent the similarity of the edge between patent ${i}$ and patent ${j}$.

	$\mathbf{D}_{ij}$ is an ${I}$-dimension column vector to measure the impact of different aspects, and the value in row ${k = \left\{ {1, \cdots ,I} \right\}}$ represents the impact of the target patent on a specific aspect. ${{\mathbf{W}_c}^T \in {\mathbb R^{I \times I}}}$ and ${{\mathbf{W}_n}^T \in {\mathbb R^{I \times I}}}$ are transformation matrices, ${\mathbf{b} \in {\mathbb R^{L \times 1}}}$ is a bias vector.
	\vspace{-2mm}
	\begin{equation}
		\\{\mathbf{D}_{ij}} = {\mathbf{W}_c}^T{\mathbf{c}_{ij}} + {\mathbf{W}_e}^T{\mathbf{e}_{ij}} + \mathbf{b}
		\label{D_ij_eq}
		\vspace{-2mm}
	\end{equation}
	
	For different aspects of impact, we hope to recognize the most important aspect of the target patent to the source patent to indicate the reason for the citation between them. To this end, ${{\mathbf{\alpha} _{ij}} \in {\mathbb R^{L \times 1}}}$ is calculated as Equation~\eqref{alpha_ij_eq}.
	\vspace{-2mm}
	\begin{equation}
		\\{\mathbf{\alpha} _{ij}}(k) = one\_hot(\mathop {{\rm{argmax}}}\limits_k [{g_k} + \log {\pi _k}])
		\label{alpha_ij_eq}
		\vspace{-2mm}
	\end{equation}
	where ${{\mathbf{\alpha} _{ij}(k)} \in {\mathbb R^{L \times 1}}}$ represents the ${k}$-th element in column vector ${{\mathbf{\alpha} _{ij}}}$. ${{\pi _k}(k = 1, \cdots ,I)}$ represents class probabilities. ${g_k}$ is from the Gumbel(0, 1) distribution. More details can be found in reference [4].
	
	Then, we can easily get:
	\vspace{-2mm}
	\begin{equation}
		\\{\mathbf{Y}_{ij}} = \max ({\mathbf{\alpha} _{ij}} \odot {\mathbf{D}_{ij}},0)
		\label{Y_ij_eq}
		\vspace{-2mm}
	\end{equation}
	
	According to the influence degrees of different aspects calculated above, the influence of each node in many aspects will also change.
	\vspace{-2mm}
	\begin{equation}
		\\\mathbf{{D^q}}(n + 1) = \mathbf{P}\mathbf{{D^q}}(n)
		\label{D_q_eq}
		\vspace{-2mm}
	\end{equation}
	where ${\mathbf{P}}$ is the projection matrix, which is obtained from the parameter update of system ${S_Y}$ learning.
	\vspace{-2mm}
	\begin{equation}
		\\\mathbf{P} = \beta \mathbf{E} + \upsilon (\mathbf{X} + \mathbf{Z})
		\label{P_eq}
		\vspace{-2mm}
	\end{equation}

	Specifically, ${\beta}$ is a hyperparameter. In this paper, let ${\beta  = 0.05/N}$, where ${N}$ is the number of nodes. ${\nu}$ is a parameter. The elements of the matrix ${\mathbf{E} \in {\mathbb R^{N \times N \times k}}}$ are all ${1}$, ${N}$ represents the number of nodes in the network . ${\mathbf{X}}$ is used to process nodes with edges in the network,  ${\mathbf{Z}}$ is a constant matrix used to process dangling nodes, and its element values are ${1/N}$. And ${n}$ is the propagation time step. Therefor, ${\mathbf{X}}$ is defined as follows:
	\vspace{-2mm}
	\begin{equation}
		\\\mathbf{X}(:,:,k) = \mathbf{x}{(k)^{i \to j}},\;\;k = 1, \cdots ,I
		\vspace{-1mm}
	\end{equation}
	
	We define ${\mathbf{x}(k) \in {\mathcal R^{N \times N}}}$ as Equation (13), which is used to indicate that node ${i}$ is connected with ${j}$, and patent ${j}$ has a direct impact on patent ${i}$ in terms of ${k}$:
	\vspace{-1.5mm}
	\begin{equation}
		\\\mathbf{x}{(k)^{i \to j}} = \left\{ \begin{gathered}
			\frac{{\mathbf{Y}{{(k)}^{i \to j}}}}{{\sum\limits_{i = 1}^N {\mathbf{Y}{{(k)}^{i \to j}}} }}\;, if\; < {v_i},{v_j},{t_{ij}} >  \in E  \; \hfill \\
			0\;\;\;\;\;\;\;\;\;\;\;\;\;\;,otherwise \hfill \\ 
		\end{gathered}  \right.
		\vspace{-1.5mm}
	\end{equation}
	
	And define ${\mathbf{Z}}$ as equation (13):
	\vspace{-1.5mm}
	\begin{equation}
		\\\mathbf{Z}(:,:,k) = \left\{ \begin{gathered}
			\frac{1}{N}\mathbf{E}(:,:,k),\;\;if\;{v_i}\;is\;dangling\;node \hfill \\
			0,\;\;otherwise \hfill \\ 
		\end{gathered}  \right.
		\vspace{-1.5mm}
	\end{equation}
	
	By learning two coupling systems ${S_Y}$ and ${S_D}$ jointly, we can learn the stable network representation of subgraphs ${G^i}$.
	\vspace{-1.5mm}
	\begin{equation}
		\\S_Y:{\mathbf{Y}^{q + 1}} = {f_1}({\mathbf{Y}^q},{\mathbf{R}^q}\left| {\tilde G} \right.)
	\end{equation}
	\vspace{-4mm}
	\begin{equation}
		\\S_D:{\mathbf{D}^{q + 1}} = {f_2}({\mathbf{Y}^{q + 1}},{\mathbf{D}^q}\left| {\tilde G} \right.)
		\vspace{-1.5mm}
	\end{equation}	
	where ${q}$ is the ${q}$-th alternate optimization, ${\mathbf{D}^q}$ is the influence state of different aspects in the ${q}$-th iteration, ${\mathbf{Y}^q}$ is the influence degree of different aspects in the ${q}$-th iteration, ${\mathbf{Y}^{q+1}}$ is the value after ${S_Y}$ learning, and is the network state after ${S_D}$ learning.
	
	Define two ${loss{_e}}$ and ${loss{_t}}$ to optimize the topological structure of the patent reference network and semantic interaction of patent text, respectively.
	\vspace{-1.5mm}
	\begin{equation}
		\\los{s_e} = \max \left\{ {0,{\lambda _e} - ({\mathbf{F}_{i,j}} - {\mathbf{F}_{i,k}})} \right\}
	\end{equation}
	\vspace{-5mm}
	\begin{equation}
		\\los{s_t} = \max \left\{ {0,{\lambda _t} - ({\mathbf{D}_{i,j}}\left| {{\alpha _{i,j}}} \right. - {\mathbf{D}_{i,k}}\left| {{\alpha _{i,j}}} \right.)} \right\}
		\vspace{-1.5mm}
	\end{equation}
	where, ${{\lambda _e}}$ and ${{\lambda _t}}$ are parameters of the loss function. It should be satisfied that, given a source node ${i}$ and any two other nodes ${j}$ and ${k}$, if there is no edge between node ${k}$ and node ${i}$, and there is an edge between node ${j}$ and node ${i}$, then ${{\mathbf{D}_{i,j}}\left| {{\alpha _{i,j}}} \right. > {\mathbf{D}_{i,k}}\left| {{\alpha _{i,j}}} \right.}$.
	
			%

	\section{Experiment}
	We implement PatSTEG using Keras, and the code is provided on an anonymous site\footnote{https://github.com/MiaoRan668/PatSTEG} for reproducibility review. In this section, we conduct extensive experiments to evaluate the proposed PatSTEG framework and other state-of-the-art methods on various patent citation prediction tasks.
	
	Patent citation prediction is a typical link prediction problem, where the objective is to forecast missing or forthcoming links in a network. 
	To examine the performance of the PatSTEG and other compared methods, we undertake a series of experiments on diverse citation datasets, including our real-world patent citation datasets. Our study includes a detailed investigation of numerous performance metrics, including top-k average precision (AP) and nDCG. Furthermore, we present a full explanation of insights garnered from our studies.
	
	\subsection{Experimental protocol}
	
	
	\textbf{Datasets}. We apply four datasets in our research, including five public datasets (Cora, PubMed, DBLPv7~\cite{moreira2015learning}, and a real-world dataset of patents in China (named CNPat). Details about public datasets are summarized in Table \ref{tab:publicdatasummary}.
	
	\begin{table}[!t]
		\centering
		\setlength\tabcolsep{4pt}
		\caption{Summary of public datasets}
		\label{tab:publicdatasummary}
		\resizebox*{\linewidth}{!}{
			\begin{tabular}{c|cccccc}
				\toprule
				Datasets & \# Nodes & \# Edges & \# Features & \# Density & \# Byte Size\\
				\midrule
				Cora & 2708 & 5429 & 1433 & 0.074 \% & 4.5 MB \\
				CiteSeer & 3312 & 4732 & 3703 & 0.030 \% & 5.9 MB \\
				PubMed & 19717 & 44338 & 500 & 0.011\% & 8.3 MB \\
				DBLPv7 & 29896 & 59549 & - & 0.007\% & 15.7 MB \\
				\bottomrule
			\end{tabular}\par
		}
	\end{table}
	
	In addition, we have collected Chinese patent information in the past 30 years from the database of Intellectual Property Publishing House. The patent citation datasets are categorized by International Patent Classification (IPC) codes, separately named from CNPatA to CNPatH. The summary of these datasets is listed as Table~\ref{tab:patentdatasummary}. 
	
	\begin{table}[!t]
		\centering
		\setlength\tabcolsep{4pt}
		\caption{Summary of patent datasets}
		\label{tab:patentdatasummary}
		\resizebox*{\linewidth}{!}{
			\begin{tabular}{c|cccc}
				\toprule
				Datasets 	& \# Nodes 	& \# Edges 	& \# Density	& Patent Subject	\\ 
				\midrule
				CNPatA		& 54352		& 61858		& 0.0023\%		& Human necessities	\\
				CNPatB		& 59773		& 68727		& 0.0019\%		& Operation, transportation	\\
				CNPatC		& 51962		& 59990		& 0.0022\%		& Chemistry, metallurgy	\\
				CNPatD		& 44848		& 53902		& 0.0027\%		& Textile, papermaking	\\
				CNPatE		& 52427		& 62618		& 0.0023\%		& Fixed constructions	\\
				CNPatF		& 51962		& 59990		& 0.0022\%		& Mechanical engineering, lighting, etc.	\\
				CNPatG		& 52156		& 48262		& 0.0018\%		& Physics	\\
				CNPatH		& 48322		& 57767		& 0.0025\% 		& Electricity	\\	
				
				\bottomrule
			\end{tabular}\par
		}
	\end{table}
	
	\textbf{Compared models}. We compare PatSTEG with the following models: Node2Vec~\cite{GroverL16}, GCN~\cite{KipfW17}, GraphSAGE~\cite{hamilton2017inductive}, SEAL~\cite{zhang2018link}, GNAE~\cite{ahn2021variational}, VGNAE~\cite{ahn2021variational}.
	
	\textbf{Metrics}. We conduct an experimental evaluation of the contrasted model by the following metrics: AUC, Average Precision@k, Recall, and nDCG.

	\begin{table*}[h]
		\centering
		\caption{Performance comparison on the public patent datasets and CNPatG dataset.}
		\setlength\tabcolsep{5pt} 
		\label{tab:result1}
		\begin{tabular}{ccccccccccccccccc}
			\hline
			\multirow{2}{*}{Dataset} &  & \multicolumn{3}{c}{Cora}                            &  & \multicolumn{3}{c}{CiteSeer}                        &  & \multicolumn{3}{c}{PubMed}                          &  & \multicolumn{3}{c}{CNPatG}                          \\ \cline{3-5} \cline{7-13} \cline{15-17} 
			&  & AUC             & AP              & Recall          &  & AUC             & AP              & Recall          &  & AUC             & AP              & Recall          &  & AUC             & AP              & Recall          \\ \hline
			Node2Vec                 &  & 0.9046          & 0.9245          & 0.7394          &  & 0.8682          & 0.8978          & 0.7707          &  & 0.6883          & 0.7550          & 0.7801          &  & 0.7044          & 0.6204          & 0.7427          \\
			GCN                      &  & 0.8240          & 0.8135          & 0.6458          &  & 0.8472          & 0.8006          & 0.7520          &  & 0.9076          & 0.8878          & 0.7201          &  & 0.7800          & 0.6440          & 0.0955          \\
			GraphSAGE                &  & 0.8975          & 0.7953          & 0.8672          &  & 0.9061          & 0.7939          & 0.9040          &  & 0.9472          & 0.8391          & 0.9249          &  & 0.5698          & 0.6415          & 0.7293          \\
			SEAL                     &  & 0.9036          & 0.9213          & 0.8387          &  & 0.9103          & 0.9253          & 0.7798          &  & 0.9079          & 0.9251          & 0.8482          &  & 0.5173          & 0.5139          & 0.5175          \\
			GNAE                     &  & 0.9418          & 0.9392          & \textbf{0.9886} &  & 0.9464          & 0.9556          & 0.9538          &  & 0.9541          & 0.9486          & \textbf{0.9888} &  & 0.6437          & 0.7946          & 0.7451          \\
			VGANE                    &  & 0.9456          & 0.9477          & 0.9744          &  & \textbf{0.9553} & \textbf{0.9578} & \textbf{0.9884} &  & 0.9524          & 0.9414          & 0.9986          &  & 0.6459          & 0.7950          & 0.7448          \\ \hline
			PatSTEG-NDP (Ours)         &  & 0.9058          & 0.8637          & 0.7516          &  & 0.8767          & 0.9048          & 0.7759          &  & 0.9587          & 0.9525          & 0.8600          &  & 0.9678          & \textbf{0.8049} & 0.7348          \\
			PatSTEG-DP (Ours)         &  & \textbf{0.9765} & \textbf{0.9712} & 0.9223          &  & 0.9285          & 0.9238          & 0.7798          &  & \textbf{0.9635} & \textbf{0.9689} & 0.8591          &  & \textbf{0.9725} & 0.8202          & \textbf{0.7506} \\ \hline
		\end{tabular}\par
	\end{table*}
	
	\subsection{Results}
	We compare the performance of PatSTEG to baselines.
	
		\textbf{The performance on patent dataset}. In the experiment, we selected our CNPatG dataset and the public DBLPv7, because there is text information of nodes in these two datasets, and we uniformly selected the title of the text as supplementary information.   From Table \ref{tab:result5}, it can be seen that both AP and nDCG achieve the best results when considering text information and dynamic learning in the two data sets of DBLPv7 and CNPatG.   On the DBLPv7 dataset, the result of AP with text information considered is about 2.5\% higher than that without text information.   On the CNPatG dataset, the result of AP with text information is about 27\% higher than that without text information.   Compared with the pre-processed public data set DBLPv7, the citation network of CNPatG is very sparse.  In real life, this sparse network is very common.   This also proves that it is better for us to consider side information in the model, which to some extent supplements the missing structural information.  
		
		In addition to text information, the dynamic learning citation network is another advantage of PatSTEG.   It can be seen that considering dynamic learning on the patent data set CNPatG will increase the AP value by about 1.5\%.   Therefore, we believe that it is a wise choice to consider both text information and dynamic learning.
		
		\textbf{The performance on public datasets}. To further verify the effect of the model, We compare PatSTEG and baselines on public datasets, including Cora, CiteSeer, and PubMed. As shown in Tabel~\ref{tab:result1}, PatSTEG-DP means PatSTEG model with dynamic propagation, while PatSTEG-NDP means PatSTEG model without dynamic propagation. It can be seen that the accuracy of PatSTEG is comparable to other models.

        \begin{figure}[!t]
		\centering
        \includegraphics[width=0.75\linewidth]{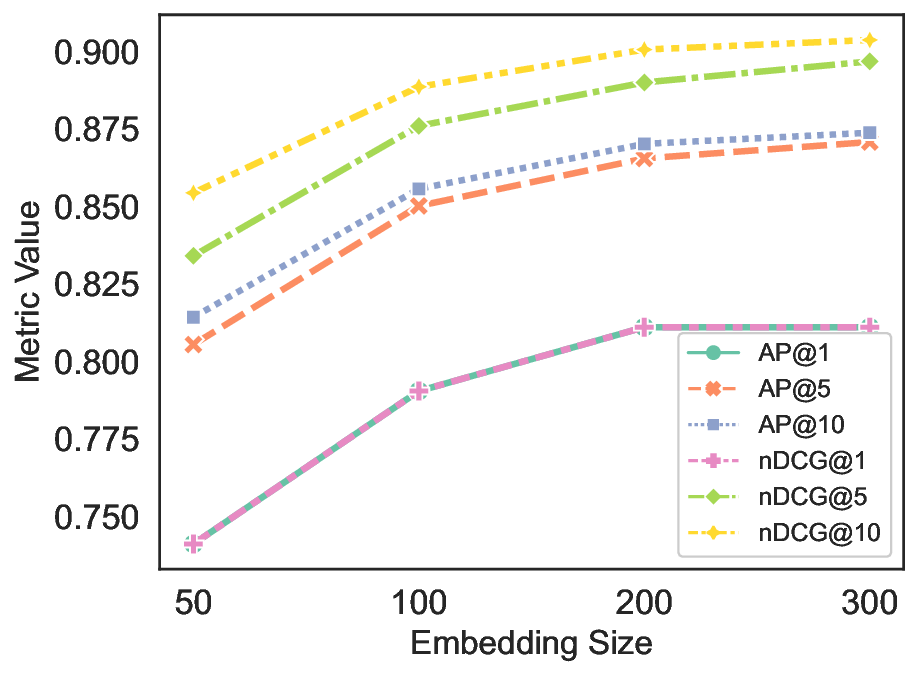}
		\caption{The performances of different embedding sizes.}
		\label{fig:compare_embed_size}
	\end{figure}
			
			\subsection{Ablation Study}
			We perform ablation studies on PatSTEG to understand the importance of different components of PatSTEG, including the text branch of PatSTEG and different embedding sizes.
				
				\textbf{Effect of embedding size}. We further investigate how embedding size has an effect on link prediction tasks in PatSTEG. In the context of machine learning-based link prediction in graphs, the embedding size refers to the dimension of vector representations learned for each node in the network. The embedding size is an essential hyperparameter that may affect the performance of link prediction algorithms. However, large embedding sizes also cost a lot for training. 
				
				Overall, the choice of embedding size is an important consideration in link prediction. Fig. ~\ref{fig:compare_embed_size} demonstrates the effect of varying embedding sizes. A noteworthy result is that an appropriate embedding size (200 in our experiments) is adequate for PatSTEG. In conclusion, the interpretability of PatSTEG applied to patent links is a vital factor that sets it distinct. The multi-aspect interactions between patent nodes enable us to identify the most relevant patents and their topics precisely. With the aid of the word clouds, we visualize the citation aspects and make a better comprehension of links between patent nodes. This characteristic will be particularly helpful for patent analysts and researchers.

 \begin{table}[!t]
			\centering
			\captionof{table}{Ablation study on PatSTEG.}
			\label{tab:result5}
			\scalebox{0.88}{
				\begin{tabular}{c ccc|ccc}
					\toprule
					\multirow{2}{*}{Text} & \multicolumn{3}{c|}{AP}                             & \multicolumn{3}{c}{nDCG}                            \\ \cmidrule{2-7}
					& @1              & @5              & @10             & @1              & @5              & @10             \\ \midrule
					title                     & 0.8092          & 0.8638          & 0.8708          & 0.8092          & 0.8916          & 0.9009          \\
					abstract                     & 0.3740          & 0.4374          & 0.4653          & 0.3740          &  0.4791          & 0.5461          \\
					claim                     & 0.3282          & 0.4145          & 0.4449          & 0.3282          &  0.4690          & 0.5398          \\
					title+abstract            & 0.8352          & \textbf{0.8876} & \textbf{0.8913} & 0.8352          & \textbf{0.9098} & \textbf{0.9178} \\
					title+claim               & 0.8068          & 0.8168          & 0.8663          & 0.8668          & 0.8852          & 0.8954          \\
					abstract+claim               & 0.3316          & 0.3956          & 0.4279          & 0.3316          & 0.4396          & 0.5152          \\
					title+abstract+claim      & \textbf{0.8358} & 0.8874          & 0.8899          & \textbf{0.8358} & 0.9065          & 0.9157          \\ 
					\bottomrule
				\end{tabular}\par}
	\end{table}

\subsection{Model Interpretability Demonstration}
\begin{figure}[!t]
	\centering
	\includegraphics[width=\linewidth]{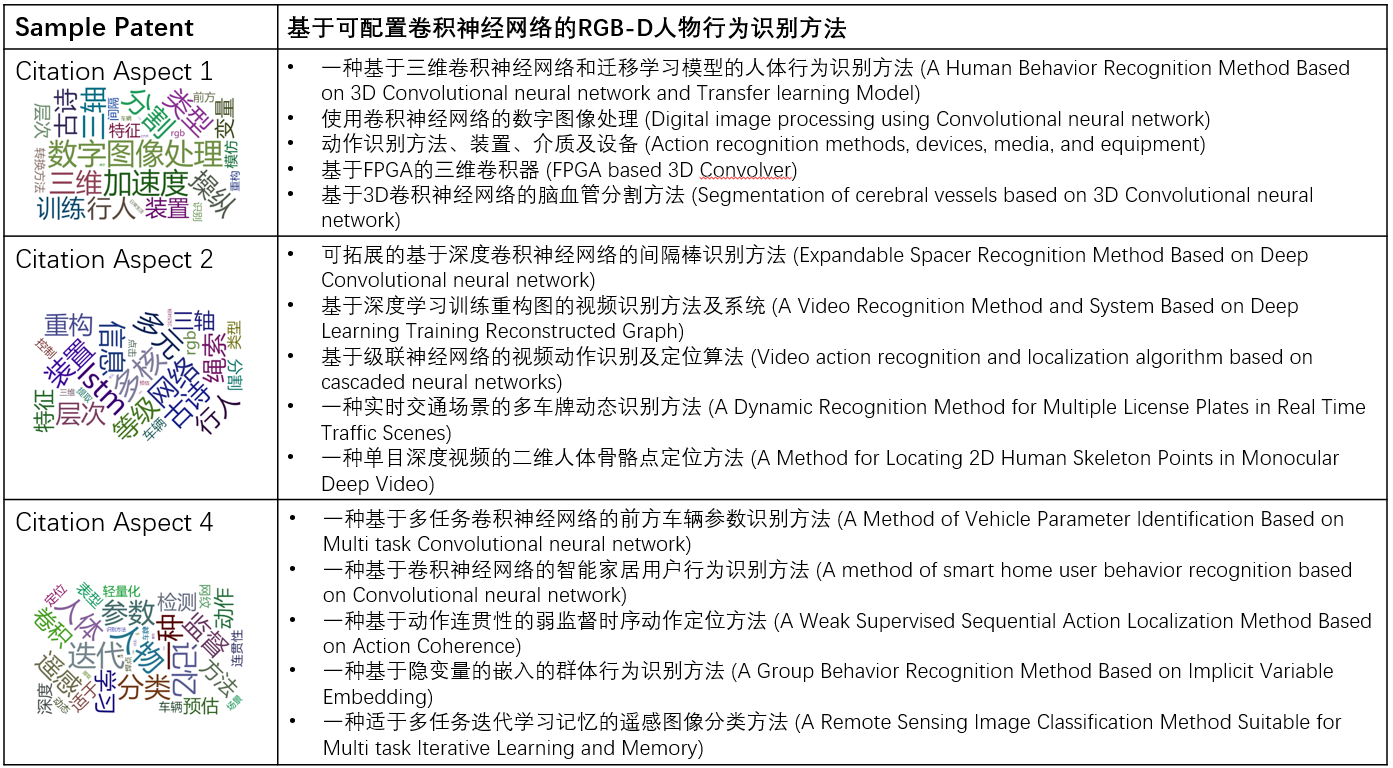}
	\caption{The different topics of citations.}
	\label{fig:explanation}
     \vspace{-3mm}
    \end{figure}

One of the most crucial elements of our approach is the interpretability of patent links. We randomly selected a patent titled ``Configurable Convolutional Neural Network-based Algorithm for RGB-D Human Behavior Detection'' from the CNPatG dataset and extracted all the patents that cite from it to illustrate the interpretability of PATCAM. We picked the top 5 patents in terms of relevance in aspects 1, 2, and 4, which are based on the interaction scores obtained by PATCAM. As illustrated in Fig~\ref{fig:explanation}, aspect 1 concentrates on convolutional neural network approaches, aspect 2 concentrates on video recognition, while aspect 4 concentrates on behavior recognition. Therefore, our technique not only predicts the citation connections between patent nodes but also gives interpretability in terms of citation links.

			\section{Conclusion}
			In this paper, we discuss the characteristics of patents in the patent database and emphasize the importance and current challenges of patent analysis. We propose a new patent citation analysis model (PatSTEG), which integrates multi-level text information. PatSTEG constructs the patent citation network based on the citation relationships among patents and leverages the textual information of patents as supplementary attributes.  Inspired by the idea of the evolutionary dynamics method, the formation process of the patent citation network is constructed and the causes of patent citation are studied. PatSTEG extracts the patents published by the Intellectual Property Publishing House as the main experimental object to explore the characteristics of the patents. In addition, the experimental results on the public benchmark data also verified the robustness and scalability of the model. It is proved that PatSTEG can be applied not only to patents with different IPCs, that is, patents from different fields, but also not be confined to a particular writing language.
			Moreover, our method can recognize different topic aspects of citations from the same patent, which can be visualized and assist in patent citation analysis.
			
			Our work provides a new idea for subsequent patent research. Further, more studies such as adding other structured information on patents are worthy of exploration.  Based on understanding the formation of patent citations, we can further predict the possible hot spots in the future. We believe that the advancements in patent citation link prediction have the potential to revolutionize patent analysis, intellectual property management, and innovation studies. By leveraging the predictive power of algorithms, we can unlock valuable insights, streamline decision-making processes, and drive technological advancements in various industries.

			\bibliographystyle{IEEEtran}
			\bibliography{reference}
			
		\end{document}